\documentclass[a4paper,fleqn,usenatbib,useAMS]{mnras}

\usepackage{graphicx}	
\usepackage{amsmath}	
\usepackage{amssymb}	

\usepackage[T1]{fontenc}
\usepackage{ae,aecompl}

\title[Broad-band spectrophotometry of HAT-P-32\,b]{Broad-band spectrophotometry of HAT-P-32\,b: Search for a scattering signature in the planetary spectrum}

\author[M. Mallonn et al.]{
M. Mallonn,$^{1}$\thanks{E-mail: mmallonn@aip.de}
I. Bernt,$^{1}$
E. Herrero,$^{2}$
S. Hoyer,$^{3,4}$
J. Kirk,$^{5}$
P. J. Wheatley,$^{5}$
\newauthor{
M. Seeliger,$^{6}$
F. Mackebrandt,$^{1}$
C. von Essen,$^{7}$
K. G. Strassmeier,$^{1}$
T. Granzer,$^{1}$
}
\newauthor{
A. K\"{u}nstler,$^{1}$
V. S. Dhillon,$^{8,3}$
T. R. Marsh,$^{5}$
and J. Gaitan$^{9}$
}
\\
\\
$^{1}$Leibniz-Institut f\"{u}r Astrophysik Potsdam (AIP), An der Sternwarte 16, D-14482 Potsdam, Germany\\
$^{2}$Institut de Ci\`{e}ncies de l'Espai (CSIC-IEEC), Campus UAB, Carrer de Can Magrans s/n, 08193 Cerdanyola del Vall\`{e}s, Spain\\
$^{3}$Instituto de Astrof\'{\i}sica de Canarias, V\'{i}a L\'{a}ctea s/n, E-38205 La Laguna, Tenerife, Spain \\
$^{4}$Universidad de La Laguna, Dpto. Astrof\'{\i}sica, E-38206 La Laguna, Tenerife, Spain \\
$^{5}$Department of Physics, University of Warwick, Gibbet Hill Road, Coventry CV4 7AL, UK \\
$^{6}$Astrophysical Institute and University Observatory Jena, Schillerg\"{a}\ss{}chen 2-3, D-07745 Jena, Germany \\
$^{7}$Stellar Astrophysics Centre (SAC), Department of Physics and Astronomy, Aarhus University, DK-8000 Aarhus C, Denmark \\
$^{8}$Department of Physics and Astronomy, University of Sheffield, Sheffield S3 7RH, UK \\
$^{9}$C / 10 Violeta, 17300 Blanes, Girona, Spain
}

\date{Accepted 2016 August 5. Received 2016 August 5; in original form 2016 May 12}

\pubyear{2016}

\begin{document}
\label{firstpage}
\pagerange{\pageref{firstpage}--\pageref{lastpage}}
\maketitle

\begin{abstract}
Multi-colour broad-band transit observations offer the opportunity to characterise the atmosphere of an extrasolar planet with small- to medium-sized telescopes. One of the most favourable targets is the hot Jupiter HAT-P-32\,b. We combined 21 new transit observations of this planet with 36 previously published light curves for a homogeneous analysis of the broad-band transmission spectrum from the Sloan u' band to the Sloan z' band. Our results rule out cloud-free planetary atmosphere models of solar metallicity. Furthermore, a discrepancy at reddest wavelengths to previously published results makes a recent tentative detection of a scattering feature less likely. Instead, the available spectral measurements of HAT-P-32\,b favour a completely flat spectrum from the near-UV to the near-IR. A plausible interpretation is a thick cloud cover at high altitudes.
\end{abstract}

\begin{keywords}
techniques: photometric -- planets and satellites: atmosphere -- planets and satellites: individual: HAT-P-32\,b
\end{keywords}



\section{Introduction}
 
Transit events offer the opportunity to characterise the atmospheres of extrasolar planets. During a transit, a fraction of the star light shines through the outmost layers of the planetary atmosphere. Depending on the opacity of its chemical constitution, the star light transmits or becomes scattered or absorbed. Therefore, the altitude of optical depth equaling unity (effective planetary radius) is a function of wavelength and depends on the atmospheric composition. The measurement of the effective planetary radius as a function of wavelength is called transmission spectroscopy and is often accomplished by spectrophotometric observations of the transit event. The observer obtains photometric transit light curves at multiple wavelengths, either simultaneously by low-resolution spectroscopic observations \citep[e.\,g.,][]{Bean2010,Gibson2013,MallonnH19} or  in different broad-band filters both simultaneously \citep[e.\,g.,][]{Nascimbeni2013,Mancini2013} or at multiple transit epochs \citep[e.\,g.,][]{
deMooij2012,Dragomir2015,MallonnH12}. 

A value of the planetary radius in relation to the (wavelength-independent) stellar radius is derived by a model fit to the transit light curve. This method was successfully applied to measure the spectroscopic absorption feature of, e.\,g., sodium, potassium, and water \citep{Sing2011b,Huitson2013,Nikolov2014}. In other target spectra the spectral features predicted by cloud-free models could be ruled out \citep[e.\,g.,][]{Gibson2013,Pont2013,MallonnH19,Lendl2016}. These spectra are either simply flat in the probed wavelength region, or show a trend of increasing opacity toward blue wavelengths explainable by Rayleigh- or Mie-scattering \citep{Jordan2013,Stevenson2014}. Interestingly, at wavelengths shorter than about 500~nm currently all gas giants on close orbits (the so-called hot Jupiters) seem to show an increase in opacity when measured with sufficient precision \citep[uncertainty of the effective planetary size about one scale height,][]{Sing2016}. 

One target of special interest is the hot Jupiter HAT-P-32\,b. It was discovered by \cite{Hartman2011} and is one of the best targets for transmission spectroscopy because of its large transit depth of more than two percent, its large planetary scale height of about 1000~km, and a relatively bright host star (V\,=\,11.4~mag). HAT-P-32\,b's atmospheric transmission spectrum lacks the predicted cloud-free absorption of Na, K and H$_2$O \citep[assuming solar composition,][hereafter G13]{GibsonH32}. This result was recently confirmed by transmission spectroscopy of \cite{Nortmann2016} (hereafter N16). However, \cite{MS2016} (hereafter Paper~I) found indications for a slope of increasing effective planetary radius toward the blue by measuring the spectrum from 330 to 1000~nm. The amplitude of this increase of only two atmospheric pressure scale heights is small compared to other hot Jupiters with measured scattering signatures, e.\,g. HD189733\,b \citep{Sing2011,Pont2013}. In this work, we attempt a verification 
of this blueward increase of effective planetary radius of HAT-P-32 by multi-epoch and multi-colour observations using broad-band photometry. 

While in principle the broad-band filters allow for low photon-noise in the photometry even with meter-sized telescopes, it has proven to be a demanding task to reach sufficient precision to discriminate between different models \citep[e.\,g.,][]{Teske2013,Fukui2013,MallonnH12}. Furthermore, for single observations per filter a potential effect of correlated noise is not always obvious \citep[e.\,g.,][]{Southworth2012}. One way to lower such potential effect is to observe multiple transit light curves per filter under the assumption that the correlated noise does not repeat because observing conditions change from night to night for ground-based observations \citep{Lendl2013,MallonnH12}. Therefore, we collected the largest sample of light curves analysed for broad-band spectrophotometry of an exoplanet so far. We present 21 new observations and analyse them homogeneously together with 36 already published light curves. Section \ref{chap_obs} gives an overview about the observations and data reduction and 
Section \ref{chap_anal} describes the analysis. The results are presented in Section \ref{chap_res} and discussed in Section \ref{chap_disc}. The conclusions follow in Section \ref{chap_concl}.

\section{Observations and data reduction}
\label{chap_obs}
Transit light curves were taken with the robotic 1.2\,m STELLA telescope, with the 2.5\,m Nordic Optical Telescope (NOT), the 0.8\,m IAC80 telescope, the 2.2\,m telescope of Calar Alto, the 70\,cm telescope of the Leibniz Institute for Astrophysics Potsdam (AIP), and the 4.2\,m William Herschel Telescope (WHT).

STELLA and its wide field imager WiFSIP \citep{Strassmeier2004,Weber2012} observed in total 11 transits in five observing seasons using the filters Johnson B and Sloan r'. The first seven transits were observed with both filters alternating. These r' band light curves were already published by \cite{Seeliger2014}. The four light curves from 2014 to 2016 were taken only in Johnson B. WiFSIP holds a back-illuminated 4k\,$\times$\,4k 15~$\mu$m pixel CCD and offers a field of view (FoV) of 22$'\times$22$'$. To minimise the read-out time, a windowing of the CCD was used reducing the FoV to about 15$'\times$15$'$. A small defocus was applied to spread the PSF to an artificial FWHM of about 3$''$.

One transit was observed with the NOT as a Fast Track program using ALFOSC in imaging mode. ALFOSC contains a 2k\,$\times$\,2k E2V CCD providing a FoV of 6.4$'\times$6.4$'$. 

One transit was observed with BUSCA, the four-channel imager at the Calar Alto Observatory 2.2\,m telescope \citep{Reif1999}. The instrument performs simultaneous photometry in four different bandpasses with a FoV of 11$'\times$11$'$. For the bandpass of shortest wavelength we used a white glass filter and the beam splitter defined the limit of $\lambda < 430$~nm. For the other bandpasses from blue to red we used a Thuan-Gunn~g, Thuan-Gunn~r, and Bessel I filter. Unfortunately, the observation of a pre-ingress baseline was lost due to weather, and the observing conditions remained to be non-photometric.
We discarded the light curve of shortest wavelength because it exhibited significantly larger correlated noise than the other three light curves.

One transit was observed with the IAC80 telescope, owned and operated by the Instituto de Astrof\'{\i}sica de Canarias (IAC), using its wide field imager CAMELOT. It is equipped with a CCD-E2V detector of 2k\,$\times$\,2k, with a pixel scale of about 0.3$''$/pixel providing a FoV of about 10.5\,$'\times$\,10.5\,$'$.

Two transits were observed with the 70\,cm telescope of the AIP, located in the city of Potsdam at the Babelsberg Observatory. The telescope is equipped with a cryogenic cooled 1k\,$\times$\,1k TEK-CCD providing a FoV of 8$'\times$8$'$. The first transit was observed in Johnson B and Johnson V quasi-simultaneously with alternating filters, the second one in Johnson V only. We applied a 3$\times$3 pixel binning to reduce the detector read-out time.

Data of an additional transit were taken with the triple beam, frame-transfer CCD camera ULTRACAM \citep{Dhillon2007} mounted at the WHT. The instrument optics allow for the simultaneous photometry in three different bandpasses. We chose the filters Sloan u', Sloan g' and a filter centred on the sodium doublet at 591.2~nm with a width (FWHM) of 31.2~nm. To avoid saturation of the brightest stars, the exposure time was extremely short with 0.3~s. However, the overheads are negligible owing to the frame-transfer technique. The detectors were read out in windowed mode. For the u' channel, ten exposures were co-added on chip before read-out. In the analysis, we used the light curves binned in time. The g' band light curve showed correlated noise of $\sim$\,3~mmag on short time scales and was excluded from the analysis.

We extended our sample of broad-band transit data by the five light curves published by \cite{Hartman2011} and 26 light curves published by \cite{Seeliger2014}. Furthermore, we searched the Exoplanet Transit Database\footnote{http://var2.astro.cz/ETD/} \citep[ETD,][]{Poddany2010} for available amateur observations of sufficient quality to be helpful in the recent work. ETD performs online a simple transit fit, and we used its derived transit depth uncertainty and a visual inspection for selecting five light curves for our purposes. The characteristics of all light curves are summarised in Table \ref{tab_overview}.

The data reduction of all new light curves except the ULTRACAM data was done as described in our previous work on HAT-P-12\,b \citep{MallonnH12}. Bias and flat-field correction was done in the standard way, with the bias value extracted from the overscan regions. We performed aperture photometry with the publicly available software SExtractor using the option of a fixed circular aperture MAG\_APER. The set of comparison stars (flux sum) was chosen to minimise the root mean square (rms) of the light curve residuals after subtraction of a second order polynomial over time plus transit model using literature transit parameter. Using the same criterion of a minimised rms, we also determined and applied the best aperture width. A significant fraction of our light curves nearly reached photon-noise limited precision. The transit light curves of 2012 Oct 12, observed simultaneously with the NOT and WHT, suffered from clouds moving through. We discarded all data points with flux levels below 50\% of its mean.

The ULTRACAM data were reduced in the same way as described in Kirk et al. (in prep), using the ULTRACAM data reduction pipeline\footnote{http://deneb.astro.warwick.ac.uk/phsaap/\\software/ultracam/html/index.html} with bias frames and flatfields used in the standard way. Again, the aperture size used to perform aperture photometry was optimised to deliver the most stable photometry. Only two useful comparison stars were observed in the FoV, and a simple flux sum of both as reference was found to give the differential light curve of lowest scatter. 

\begin{figure*}
   \centering
   \includegraphics[width=14.5cm,height=18cm]{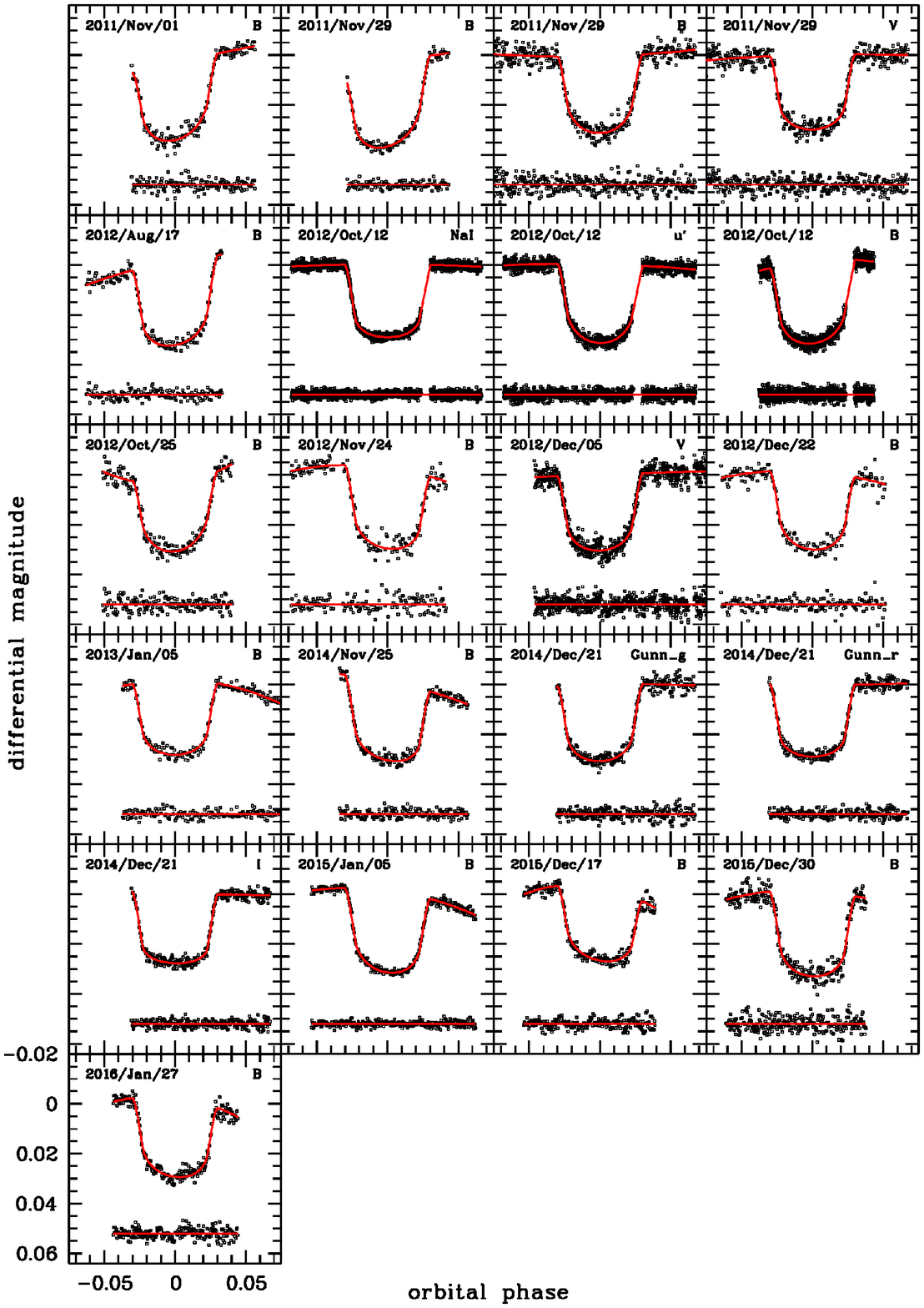}
       \caption{All light curves newly presented in this work. The red solid line denotes the individual best fit model, the resulting parameter values are given in Table \ref{tab_indiv}. Below the light curves, their corresponding residuals are shown. The scale of all panels is identical with the tickmarks labeled in the lower left.}
         \label{plot_lcs}
   \end{figure*}

The 21 new light curves are shown in Figure \ref{plot_lcs}, and the 36 re-analysed literature light curves are presented in Figure \ref{plot_lc_lit}.

\begin{table*}
\small
\caption{Overview of the analysed transit observations of HAT-P-32\,b. The columns give the observing date, the telescope used, the chosen filter, the airmass range of the observation, the exposure time, the observing cadence, the number of observed data points in-transit, the number of observed data points out-of-transit, the dispersion of the data points as root mean square (rms) of the observations after subtracting a transit model and a detrending function, the $\beta$ factor (see Section \ref{chap_anal}), and the reference of the light curves.}
\label{tab_overview}
\begin{center}
\begin{tabular}{p{19mm}p{24mm}p{10mm}p{9mm}p{9mm}p{9mm}p{9mm}p{8mm}p{8mm}l}
\hline
\hline
\noalign{\smallskip}

Date        &   Telescope   &   Filter  &   $t_{\mathrm{exp}}$ (s) &  Cadence (s) & $N_{\mathrm{it}}$ & $N_{\mathrm{oot}}$ & rms (mmag)  &  $\beta$ & Reference \\
\hline
\noalign{\smallskip}

2007 Sep 24 &   FLWO 1.2\,m        & z'     &           &   33      & 330  & 273  &   1.93 &  1.13 &  \cite{Hartman2011}  \\
2007 Oct 22 &   FLWO 1.2\,m        & z'     &           &   53      & 205  & 284  &   2.51 &  1.00 &  \cite{Hartman2011}  \\
2007 Nov 06 &   FLWO 1.2\,m        & z'     &           &   28      & 386  & 373  &   2.06 &  1.00 &  \cite{Hartman2011}  \\
2007 Nov 19 &   FLWO 1.2\,m        & z'     &           &   38      & 292  & 373  &   1.68 &  1.27 &  \cite{Hartman2011}  \\
2007 Dec 04 &   FLWO 1.2\,m        & g'     &           &   33      & 322  & 274  &   1.85 &  1.00 &  \cite{Hartman2011}  \\                                            
2011 Nov 01 &   STELLA             & B      &    30     &   115     & 95   & 45   &   1.96 &  1.04 &  this work           \\
2011 Nov 01 &   STELLA             & r'     &    15     &   115     & 92   & 41   &   1.90 &  1.00 &  \cite{Seeliger2014} \\
2011 Nov 14 &   Jena 0.6/0.9\,m    & R      &    40     &   63      & 174  & 126  &   1.76 &  1.04 &  \cite{Seeliger2014} \\
2011 Nov 29 &   STELLA             & B      &    30     &   115     & 94   & 22   &   1.21 &  1.00 &  this work           \\
2011 Nov 29 &   STELLA             & r'     &    15     &   115     & 95   & 21   &   1.30 &  1.01 &  \cite{Seeliger2014} \\
2011 Nov 29 &   Babelsberg 70\,cm  & B      &    30     &   75      & 149  & 224  &   2.65 &  1.00 &  this work           \\
2011 Nov 29 &   Babelsberg 70\,cm  & V      &    30     &   75      & 151  & 222  &   2.44 &  1.38 &  this work           \\
2011 Nov 29 &   Rozhen 2.0\,m      & V      &    20     &   23      & 461  & 172  &   0.99 &  1.00 &  \cite{Seeliger2014} \\
2011 Dec 01 &   Rozhen 2.0\,m      & R      &    20     &   23      & 470  & 218  &   1.38 &  1.99 &  \cite{Seeliger2014} \\
2011 Dec 14 &   Rozhen 0.6\,m      & R      &    60     &   63      & 177  & 78   &   1.87 &  1.09 &  \cite{Seeliger2014} \\
2011 Dec 27 &   Rozhen 2.0\,m      & R      &    20     &   39      & 280  & 66   &   1.05 &  1.81 &  \cite{Seeliger2014} \\
2012 Jan 15 &   Swarthmore 0.6\,m  & R      &    50     &   59      & 184  & 169  &   3.21 &  1.00 &  \cite{Seeliger2014} \\
2012 Aug 15 &   Rozhen 2.0\,m      & R      &    25     &   44      & 252  & 93   &   1.24 &  1.00 &  \cite{Seeliger2014} \\
2012 Aug 17 &   STELLA             & B      &    30     &   115     & 97   & 56   &   1.61 &  1.48 &  this work           \\
2012 Aug 17 &   STELLA             & r'     &    15     &   115     & 96   & 58   &   1.72 &  1.00 &  \cite{Seeliger2014} \\
2012 Sep 12 &   OSN 1.5\,m         & R      &    30     &   39      & 269  & 63   &   3.58 &  1.09 &  \cite{Seeliger2014} \\
2012 Sep 12 &   Trebur 1.2\,m      & R      &    50     &   59      & 182  & 84   &   2.02 &  1.23 &  \cite{Seeliger2014} \\
2012 Sep 14 &   OSN 1.5\,m         & R      &    30     &   40      & 278  & 181  &   1.23 &  1.58 &  \cite{Seeliger2014} \\
2012 Oct 12 &   WHT                & NaI    &    14.7   &   14.7    & 678  & 928  &   0.99 &  1.77 &  this work           \\
2012 Oct 12 &   WHT                & u'     &    19.5   &   19.5    & 509  & 712  &   1.39 &  1.58 &  this work           \\
2012 Oct 12 &   NOT                & B      &    7.0    &   14.9    & 657  & 266  &   1.82 &  1.00 &  this work           \\
2012 Oct 25 &   STELLA             & B      &    25     &   105     & 104  & 55   &   2.35 &  1.00 &  this work           \\
2012 Oct 25 &   STELLA             & r'     &    25     &   105     & 105  & 54   &   2.48 &  1.03 &  \cite{Seeliger2014} \\
2012 Oct 31 &   SON 0.4\,m         & R      &           &   42      & 261  & 150  &   3.07 &  1.00 &  ETD, P. Kehusmaa    \\
2012 Nov 07 &   0.25\,m            & V      &    180    &   190     & 53   & 24   &   2.23 &  1.00 &  ETD, J. Gaitan      \\
2012 Nov 22 &   OSN 1.5\,m         & R      &    30     &   35      & 317  & 193  &   1.05 &  1.14 &  \cite{Seeliger2014} \\
2012 Nov 24 &   STELLA             & B      &    40     &   136     & 79   & 60   &   2.57 &  1.00 &  this work           \\
2012 Nov 24 &   STELLA             & r'     &    25     &   136     & 77   & 62   &   2.60 &  1.00 &  \cite{Seeliger2014} \\
2012 Dec 05 &   Babelsberg 70\,cm  & V      &    30     &   36      & 315  & 303  &   2.41 &  1.53 &  this work           \\
2012 Dec 22 &   STELLA             & B      &    40     &   136     & 81   & 76   &   2.22 &  1.00 &  this work           \\
2012 Dec 22 &   STELLA             & r'     &    25     &   136     & 82   & 72   &   1.76 &  1.13 &  \cite{Seeliger2014} \\
2013 Jan 05 &   STELLA             & B      &    40     &   136     & 81   & 77   &   1.48 &  1.23 &  this work           \\
2013 Jan 05 &   STELLA             & r'     &    25     &   136     & 82   & 76   &   1.32 &  1.12 &  \cite{Seeliger2014} \\
2013 Aug 22 &   0.35\,m            & R      &    70     &   87      & 127  & 84   &   2.03 &  1.04 &  ETD, V.P. Hentunen  \\
2013 Sep 06 &   Rozhen 2.0\,m      & R      &    30     &   33      & 334  & 55   &   0.87 &  1.34 &  \cite{Seeliger2014} \\
2013 Sep 06 &   Jena 0.6/0.9\,m    & R      &    40     &   58      & 191  & 124  &   1.49 &  1.03 &  \cite{Seeliger2014} \\
2013 Sep 06 &   Torun 0.6\,m       & R      &    10     &   13      & 829  & 580  &   3.27 &  1.65 &  \cite{Seeliger2014} \\
2013 Oct 06 &   OSN 1.5\,m         & R      &    30     &   32      & 348  & 236  &   0.93 &  1.00 &  \cite{Seeliger2014} \\
2013 Nov 01 &   Rozhen 2.0\,m      & R      &    25     &   44      & 249  & 60   &   0.73 &  1.07 &  \cite{Seeliger2014} \\
2013 Nov 03 &   OSN 1.5\,m         & R      &    30     &   35      & 316  & 195  &   1.20 &  1.29 &  \cite{Seeliger2014} \\
2013 Dec 01 &   OSN 1.5\,m         & R      &    30     &   35      & 315  & 218  &   1.64 &  1.69 &  \cite{Seeliger2014} \\
2013 Dec 14 &   0.30\,m            & V      &           &   80      & 135  & 173  &   4.01 &  1.17 &  ETD, R. Naves       \\
2013 Dec 29 &   Trebur 1.2\,m      & R      &    50     &   58      & 187  & 105  &   2.40 &  1.00 &  \cite{Seeliger2014} \\
2014 Nov 25 &   STELLA             & B      &    60     &   94      & 117  & 61   &   1.44 &  1.25 &  this work           \\
2014 Dec 21 &   Calar Alto 2.2\,m  & Gunn g &    20     &   65      & 157  & 106  &   1.84 &  1.05 &  this work           \\
2014 Dec 21 &   Calar Alto 2.2\,m  & Gunn r &    20     &   65      & 160  & 107  &   1.39 &  1.10 &  this work           \\
2014 Dec 21 &   Calar Alto 2.2\,m  & I      &    20     &   65      & 160  & 107  &   1.33 &  1.00 &  this work           \\
2014 Dec 21 &   0.30\,m            & V      &           &   84      & 130  & 77   &   2.90 &  1.00 &  ETD, F. Campos      \\
2015 Jan 05 &   STELLA             & B      &    45     &   79      & 142  & 130  &   0.98 &  1.23 &  this work           \\
2015 Dec 17 &   STELLA             & B      &    45     &   79      & 139  & 77   &   1.73 &  1.47 &  this work           \\
2015 Dec 30 &   IAC80              & B      &    60     &   68      & 163  & 103  &   2.59 &  1.18 &  this work           \\
2016 Jan 27 &   STELLA             & B      &    45     &   79      & 142  & 67   &   1.89 &  1.56 &  this work           \\

\hline                                                                                                     
\end{tabular}
\end{center}
\end{table*}


\section{Light curve analysis}
\label{chap_anal}

In this work, we model all transit light curves with the publicly available software \mbox{JKTEBOP}\footnote{http://www.astro.keele.ac.uk/jkt/codes/jktebop.html} \citep{Southworth04,Southworth08} in version 34. It allows for a simultaneous fit of the transit model and a detrending function. Throughout this analysis, we use a second-order polynomial over time to detrend the individual light curves. The transit fit parameters consist of the sum of the fractional planetary and stellar radius, $r_{\star} + r_p$, and their ratio $k=r_p/r_{\star}$, the orbital inclination $i$, the transit midtime $T_0$, the host-star limb-darkening coefficients (LDC) $u$ and $v$ of the quadratic limb darkening law, and the coefficients $c_{0,1,2}$ of the polynomial over time. The index ``$\star$'' refers to the host star and ``p'' refers to the planet. The dimensionless fractional radius is the absolute radius in units of the orbital semi-major axis $a$, $r_{\star} = R_{\star}/a$, and $r_p = R_p/a$. The planetary eccentricity 
is fixed to zero following \cite{Zhao2014} and the orbital period $P_{\mathrm{orb}}$ to 2.15000825 days according to \cite{Seeliger2014}.

We modeled the stellar limb darkening with the quadratic, two-parameter law. Theoretical values for the LDC have been obtained by \cite{Claret2013} using the stellar parameter from \cite{Hartman2011} for the case of a circular planetary orbit. Following \cite{Southworth08} and our previous work on HAT-P-32\,b in Paper~I, we fitted for the linear LDC $u$ and kept the quadratic LDC $v$ fixed to its theoretical values, perturbing it by $\pm$0.1 on a flat distribution during the error estimation.

\begin{figure*}
   \centering
   \includegraphics[height=16cm,width=10cm,angle=270]{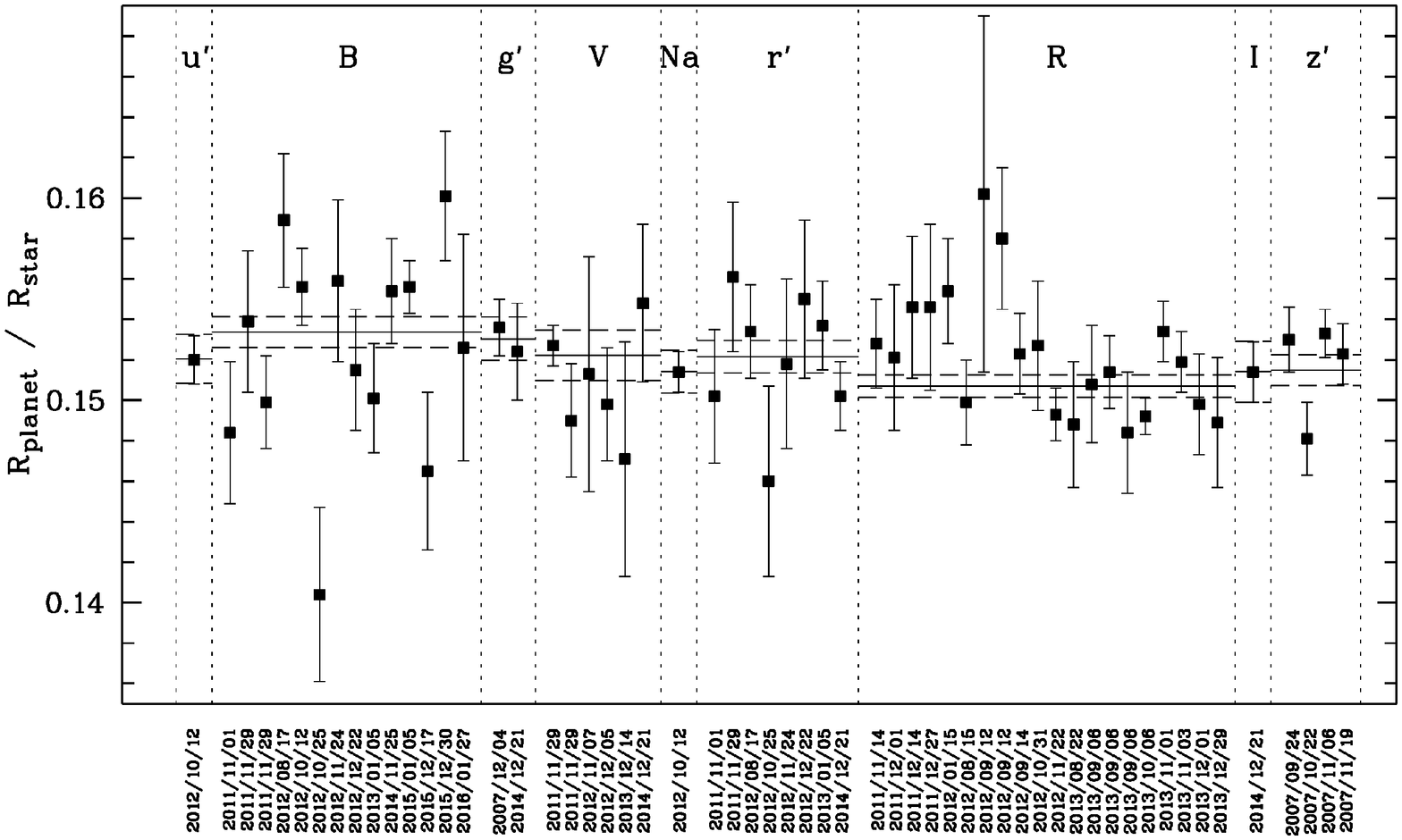}
      \caption{Planet-star radius ratio $k$ for the 57 individual transit light curves in the same order as in Table \ref{tab_indiv}. The horizontal lines mark the $k$ value for the joint fit of all light curves per filter with their uncertainties given by the horizontal dashed lines.}
         \label{plot_ind}
   \end{figure*}

Following the same procedure as in \cite{MallonnH12}, we begin the analysis of each light curve with an initial fit followed by a 3.5\,$\sigma$ rejection of outliers. After a new transit model fit, we re-scale the photometric uncertainties derived by SExtractor to yield a reduced $\chi^2$ of unity for the light curve residuals. Furthermore, we calculate the so-called $\beta$ factor, a concept introduced by \cite{Gillon06} and \cite{Winn08} to include the contribution of correlated noise in the light curve analysis. It describes the evolution of the standard deviation $\sigma $ of the light curve residuals when they become binned in comparison to Poisson noise. In the presence of correlated noise, $\sigma $ of the binned residuals is larger by the factor $\beta$ than with pure uncorrelated (white) noise. The value used here for each light curve is the average of the values for a binning from 10 to 30 minutes in 2 minute steps. In the final transit fit, we fix the $i$ and $r_{\star}$ to the 
values used in G13 and Paper~I to achieve directly comparable values of $k$. We assume their uncertainties to be a common source of noise to all bandpasses, negligible in the search for relative variations of $k$ over wavelength. Therefore, the derived errors on $k$ are relative uncertainties. The free parameters per light curve have been $r_p$, $u$, and $c_{0,1,2}$. For the bandpasses with more than one light curve, we perform an individual fit per light curve, summarised in Table \ref{tab_indiv}, and a joint run of all light curves per bandpass fitted simultaneously, summarised in Table \ref{tab_joint}. In the joint fit, the free parameters are $r_p$, $u$, and a set of detrending coefficients per involved light curve. Here, we added the Thuan-Gunn~r light curve to the sample of r' data, and Thuan-Gunn~g to the g' data because of significant overlap in the filter transmission curves. The $k$ values of the individual and the joint fits are shown in Figure \ref{plot_ind}.

The estimation of the transit parameter uncertainties was done with  ``task 8'' in \mbox{JKTEBOP} \citep{Southworth2005}, which is a Monte Carlo simulation, and with ``task 9'' \citep{Southworth08}, which is a residual-permutation algorithm that takes correlated noise into account. We run the Monte Carlo simulation with 5000 steps. As final parameter uncertainties we adopted the larger value of both methods.

The time stamps from all light curves analysed in this work were transferred to BJD$_{\mathrm{TDB}}$ following the recommendation of \cite{Eastman2010}. All individually derived transit midtimes of the light curves of our sample are in agreement with the ephemeris of \cite{Seeliger2014}.

\cite{Adams2013} found a companion object only 2.9\,$''$ to HAT-P-32, which was classified as a mid-M dwarf by \cite{Zhao2014}. The flux of this object is fully included in the aperture used for the aperture photometry for all our light curves and dilutes the transit depth as third light contribution to the star-planet system. All values of $k$ derived from r', R, I, and z' band light curves were corrected using Eq.~4 in \cite{Sing2011} and the third light as a function of wavelength measured in Paper~I.

In Paper~I we also measured and analysed the photometric variability of the host star and concluded that HAT-P-32 is photometrically constant without significant influence of stellar activity. Therefore, we assume there is no time dependence of $k$ in the interval of our observations. We computed the reduced $\chi^2$ value of $k$ of the individual transits per filter versus the $k$ value of the joint fit per filter and obtain the values of 4.2, 1.5, 1.7, and 2.4 for the bandpasses B, V, r', and R, respectively. These values are larger than unity and indicate underestimated individual error bars potentially caused by undetected correlated noise \citep{Southworth2009,Southworth2011}. Therefore, we conservatively inflate the uncertainties of $k$ of all individual light curves by the factor 1.5, which leads to an average reduced $\chi^2$ of unity. 
The error bars of $k$ of the joint fits per filter are in all cases the outcome of the residual-permutation algorithm, whose value decreases more conservatively with increasing number of light curves per joint fit than the uncertainty value derived by the Monte Carlo simulation.

\section{Results}
\label{chap_res}

   \begin{figure}
   \centering
   \includegraphics[height=\hsize,angle=270]{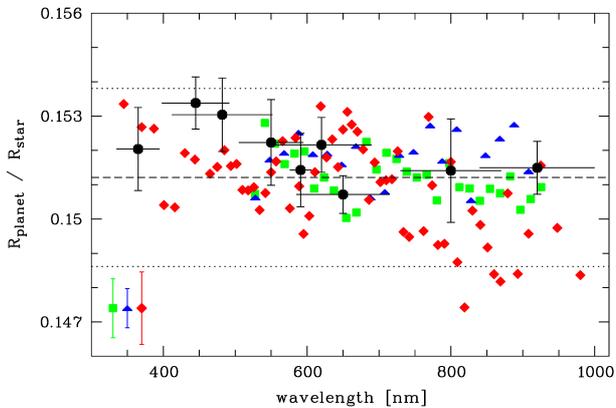}
      \caption{Broad-band transmission spectrum of HAT-P-32\,b. This work is shown in black circles, horizontal bars indicate the width of the corresponding filter curve. Spectra published by G13 (green rectangular), N16 (blue triangle), and Paper~I (red diamond) are shown for comparison. The average uncertainties of G13, N16, and Paper~I are presented in the lower left. The dotted horizontal lines indicate plus and minus two scale heights of the mean value.}
         \label{plot_1pan}
   \end{figure}

\subsection{Comparison to published spectra of HAT-P-32\,b}
The broad-band transmission spectrum of this work is shown in Figure \ref{plot_1pan} together with previously published data of HAT-P-32\,b. The results presented here are in very good agreement to the transmission spectra derived by G13 and N16 (note that N16 uses slightly different values for $i$ and $r_{\star}$). There are small discrepancies to the results of Paper~I. The B band measurement differs by about 2\,$\sigma$, and the R band value deviates by 2.5\,$\sigma$, whereas at all other wavelength there is an agreement within 1\,$\sigma$. All four investigations, G13, N16, Paper~I, and this work show good consistency shortwards of $\sim$\,720~nm. At redder wavelengths of the Cousins I and Sloan z' band, Paper~I deviates from the other three in exhibiting lower values of $k$ of about one scale height. We experimented again with the LDC by fixing $u$ and $v$ in the analysis to the values derived in the three previous publications and found that the deviation of $k$ in Paper~I cannot be explained by the 
slight discrepancies in the LDC. Instead, a potential origin of the deviations are systematics in the light curves of Paper~I.

\subsection{Treatment of stellar limb darkening}
The treatment of limb darkening in the modelling of exoplanet transit light curves is a debated topic. Multiple stellar limb darkening laws exist in varying complexity with the two-parameter, quadratic law as the one with broadest application. The limb darkening coefficients $u$ and $v$ can be either included in the fitting process as free parameters \citep[as suggested by, e.\,g.,][]{Csizmadia2013} or fixed to theoretical values  \citep[e.\,g.,][]{Claret2013}. Both options include advantages and disadvantages \citep[see, e.\,g.,][]{Mueller2013}. Therefore, \cite{Southworth08} introduced an intermediate solution by fitting for the linear LDC $u$ while keeping the quadratic LDC $v$ fixed to theoretical values for the quadratic limb darkening law. We used this option here and in Paper~I, finding in both a significant discrepancy between the fitted value of $u$ and its theoretical value from \cite{Claret2013}. For the majority of transiting exoplanet host stars, the measured LDC agree reasonably well to their 
theoretical predictions \citep{Mueller2013}. However, another known exception is HD\,209458 \citep{Knutson2007,Claret2009}. 

\begin{figure}
   \centering
   \includegraphics[width=\hsize]{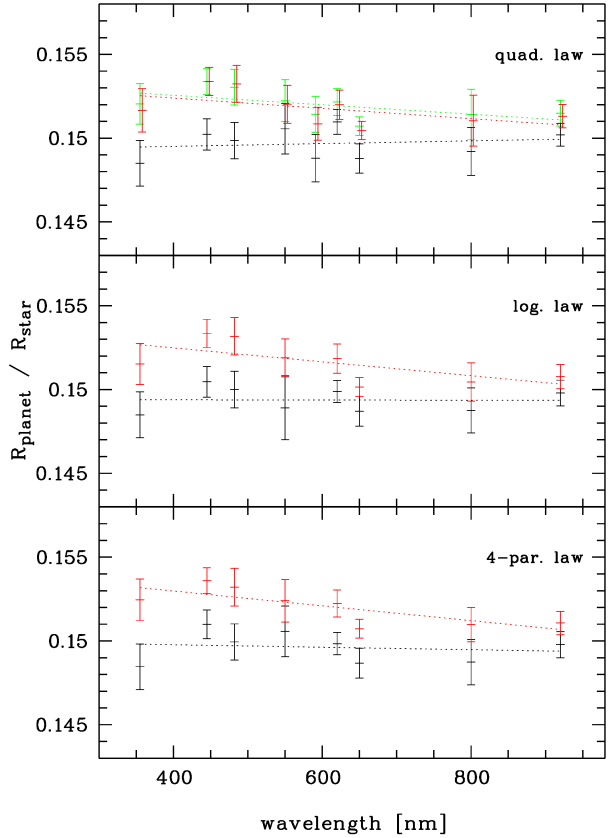}
      \caption{Planet-star radius ratio over wavelength of HAT-P-32\,b derived by the usage of different stellar limb darkening laws. Different colours denote different treatment of the LDC. Upper panel: quadratic stellar limb darkening law. Black shows the $k$ values for both LDC fixed to theoretical values, red denotes both LDC fitted, and green linear LDC fitted, but quadratic LDC fixed. Middle panel: logarithmic stellar limb darkening law. Black denotes both LDC fixed, red implies both LDC fitted. Lower panel: four-parameter stellar limb darkening law. Black denotes all four LDC fixed, red implies first LDC fitted, the other three fixed. All dotted lines show a linear regression corresponding to the data points of same colour. }
         \label{plot_ldc}
   \end{figure}

We repeated our analysis of the light curves of HAT-P-32\,b by \textit{i)} keeping both LDC fixed to theoretical values of \cite{Claret2013} and \textit{ii)} fitting for both $u$ and $v$ in the light curve fit. The results are shown in Figure \ref{plot_ldc}. While option i) results in a poorer fit and a general offset in planet-star radius ratio $k$ towards lower values, there is a very good agreement when both $u$ and $v$ were free-to-fit compared to $u$ free and $v$ fix. However, the relative variation of $k$ over wavelength remains nearly the same, irrespective of the treatment of the limb darkening. We verified this result further by the usage of the logarithmic limb darkening law suggested by \cite{Espinoza2016} for the stellar temperature of HAT-P-32 and by the usage of the four-parameter law introduced by \cite{Claret2000}. 
The resulting values of $k$ over wavelength remain nearly unaffected by the choice of the limb darkening law, the same holds for the discrepancy when the LDC are free or fixed in the fit. Nevertheless, the treatment of the LDC affects the spectral slope. A fit of a linear regression results in a slight increase of $k$ towards shorter wavelengths for free LDC, while the slope is nearly zero in the versions of fixed LDC (Figure \ref{plot_ldc}). However, in agreement to G13, N16, and Paper~I, we consider the version of free LDC more reliable because of a better fit for all bandpasses. We note that using the theoretical LDC would reduce the agreement of the data with atmosphere models containing an enhanced opacity at short wavelengths (see Section \ref{chap_theo_mod}). Therefore, it would not contradict the outcome of this work. We continue with the results obtained with free LDC.


\subsection{Comparison to theoretical models}
\label{chap_theo_mod}
\begin{figure}
   \centering
   \includegraphics[height=\hsize,angle=270]{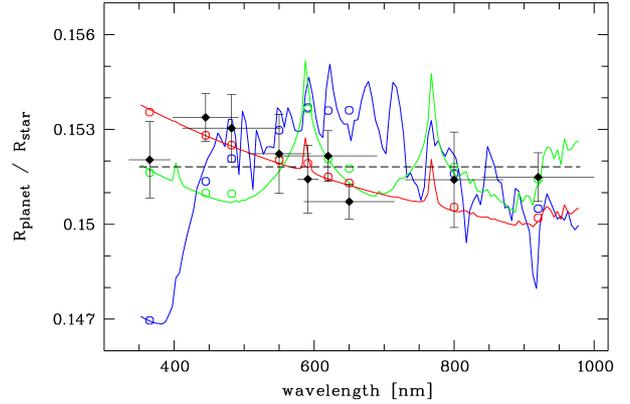}
      \caption{Broad-band transmission spectrum of HAT-P-32\,b. The measured values of this work are given in black, horizontal bars show the width of the corresponding filter curve. Overplotted are a cloud-free solar-composition model of 1750~K (blue line), a cloud-free solar composition model of 1750~K without TiO (green line), and a solar composition model of 1750~K including H$_2$ Rayleigh scattering increased by a factor of 100 \citep[red line,][]{Fortney2010}.The colour-coded open circles show the bandpass-integrated theoretical values.}
         \label{plot_spec}
   \end{figure}

We compared the derived spectrophotometric transmission spectrum of this work with theoretical models that were supplied by \cite{Fortney2010} and calculated for HAT-P-32\,b (see Fig. \ref{plot_spec}). The only fitted parameter is a vertical offset. For each model we calculated the $\chi^2$ value and the probability $P$ of the $\chi^2$ test, i.\,e. the probability that the measurements could result by chance if the model represented the true planetary spectrum. The values are summarised in Table \ref{tab_chi2}. A cloud-free solar-composition model dominated by TiO absorption and a solar-composition model with TiO artificially removed are ruled out by the nine data points of this work, in agreement to results of G13, N16 and Paper~I. Looking specifically at the theoretically predicted, prominent sodium D-line, the data point from the ULTRACAM filter shows no extra absorption in agreement to the three previous studies. A model of solar-composition without TiO including a Rayleigh scattering component with a 
cross section 100$\times$ that of molecular hydrogen yields $\chi^2 = 7.3$ for eight degrees of freedom (DOF), indicating good agreement between model and data. Similarly, a wavelength-independent planet-star radius ratio (flat spectrum) also yields an acceptable fit. Therefore, the measurements of this work cannot distinguish between the Rayleigh model and the flat model. 

We attempt a combined comparison of the results derived here and in Paper~I to the theoretical models, justified by the homogeneity of both analyses. Because of the deviating red data points of Paper~I mentioned above, we restrict the wavelength range of these data to $\lambda<720$~nm. The values of $\chi^2$ and $P$ are given in Table \ref{tab_chi2}. This combined data set shows only a very low probability for a Rayleigh slope of 0.017, whereas a flat line is significantly favoured. The low-amplitude scattering slope proposed in Paper~I also yields a good fit ($\chi^2=50.7$, $\mathrm{DOF} = 48$), though of lower quality than the flat line.

For another test, we merge the measurements of all four studies G13, N16, Paper~I, and this work, and compute a regression line. For simplicity and to account for inhomogeneities in the analyses, all data points were given equal weight. The resulting slope is $(-3.28\,\pm\,0.65)\,\times\,10^{-6}$~nm$^{-1}$. However, excluding the data points $\lambda>720$~nm of Paper~I results in a weaker, insignificant slope of $(-1.16\,\pm\,0.60)\,\times\,10^{-6}$~nm$^{-1}$, which means a consistency of the planet-star radius ratio between 350 and 1000~nm to within half of a scale height. A homogeneous re-analysis of all available data sets of HAT-P-32\,b would be needed to reliably explore the potential slope at this sub-scale height precision.


\begin{table*}
\small
\caption{Planet-to-star radius ratio $k$ per observation with relative uncertainties. In Column 4, $k$ is derived with $u$ as free parameter, with the $u$ value given in Column 5. The LDC $v$ is fixed in the analysis to its theoretical value given in Column 7. The theoretical of $u$ is given in Column 6 for comparison. Column 8 gives the applied third-light correction of $k$.}
\label{tab_indiv}
\begin{center}
\begin{tabular}{p{19mm}p{21mm}p{11mm}lcccl}
Date        &   Telescope   &   Filter  &    $k$  &  $u_{\mathrm{fit}}$  & $u_{\mathrm{theo}}$  & $v_{\mathrm{theo}}$  &  $\Delta k_{\mathrm{third\ light}}$ \\
\hline
2012 Oct 12 &   WHT              & u'    &  0.1520 $\pm$ 0.0012  & 0.543 $\pm$ 0.023   &   0.696 &  0.112  & 0         \\
\hline
2011 Nov 01 &   STELLA           & B     &  0.1484 $\pm$ 0.0035 & 0.559 $\pm$ 0.038   &   0.583 &  0.208  & 0         \\
2011 Nov 29 &   STELLA           & B     &  0.1539 $\pm$ 0.0035 & 0.512 $\pm$ 0.017   &   0.583 &  0.208  & 0         \\
2011 Nov 29 &   Babelsberg       & B     &  0.1499 $\pm$ 0.0023 & 0.495 $\pm$ 0.022   &   0.583 &  0.208  & 0         \\
2012 Aug 17 &   STELLA           & B     &  0.1589 $^{+\,\,0.0033}_{-\,0.0035}$ & 0.438 $\pm$ 0.058   &   0.583 &  0.208  & 0         \\
2012 Oct 12 &   NOT              & B     &  0.1556 $\pm$ 0.0019 & 0.486 $\pm$ 0.014   &   0.583 &  0.208  & 0         \\
2012 Oct 25 &   STELLA           & B     &  0.1404 $\pm$ 0.0043 & 0.535 $\pm$ 0.020   &   0.583 &  0.208  & 0         \\
2012 Nov 24 &   STELLA           & B     &  0.1559 $\pm$ 0.0040 & 0.435 $\pm$ 0.059   &   0.583 &  0.208  & 0         \\
2012 Dec 22 &   STELLA           & B     &  0.1515 $^{+\,\,0.0031}_{-\,0.0029}$ & 0.479 $\pm$ 0.041   &   0.583 &  0.208  & 0         \\
2013 Jan 05 &   STELLA           & B     &  0.1501 $^{+\,\,0.0026}_{-\,0.0028}$ & 0.430 $\pm$ 0.053   &   0.583 &  0.208  & 0         \\
2014 Nov 25 &   STELLA           & B     &  0.1554 $^{+\,\,0.0026}_{-\,0.0025}$ & 0.400 $\pm$ 0.040   &   0.583 &  0.208  & 0         \\
2015 Jan 05 &   STELLA           & B     &  0.1556 $\pm$ 0.0013 & 0.484 $\pm$ 0.017   &   0.583 &  0.208  & 0         \\
2015 Dec 17 &   STELLA           & B     &  0.1465 $\pm$ 0.0039 & 0.514 $\pm$ 0.051   &   0.583 &  0.208  & 0         \\
2015 Dec 30 &   IAC80            & B     &  0.1601 $\pm$ 0.0032 & 0.440 $\pm$ 0.037   &   0.583 &  0.208  & 0         \\
2016 Jan 27 &   STELLA           & B     &  0.1526 $\pm$ 0.0056 & 0.492 $\pm$ 0.063   &   0.583 &  0.208  & 0         \\
\hline
2007 Dec 04 &   FLWO 1.2\,m      & g'    &  0.1536 $^{+\,\,0.0015}_{-\,0.0014}$ & 0.465 $\pm$ 0.018   &   0.545 &  0.205  & 0         \\
2014 Dec 21 &   Calar Alto       & Gunn g&  0.1524 $^{+\,\,0.0025}_{-\,0.0023}$ & 0.400 $\pm$ 0.021   &   0.518 &  0.221  & 0         \\
\hline
2011 Nov 29 &   Rozhen 2.0\,m    & V     &  0.1527 $\pm$ 0.0010 & 0.395 $\pm$ 0.014   &   0.458 &  0.229  & 0         \\
2011 Nov 29 &   Babelsberg       & V     &  0.1490 $^{+\,\,0.0029}_{-\,0.0027}$ & 0.463 $\pm$ 0.035   &   0.458 &  0.229  & 0         \\
2012 Nov 07 &   0.25\,m          & V     &  0.1513 $\pm$ 0.0058 & 0.397 $\pm$ 0.046   &   0.458 &  0.229  & 0         \\
2012 Dec 05 &   Babelsberg       & V     &  0.1498 $\pm$ 0.0028 & 0.476 $\pm$ 0.049   &   0.458 &  0.229  & 0         \\
2013 Dec 14 &   0.30\,m          & V     &  0.1471 $\pm$ 0.0058 & 0.240 $\pm$ 0.070   &   0.458 &  0.229  & 0         \\
2014 Dec 21 &   0.30\,m          & V     &  0.1548 $^{+\,\,0.0042}_{-\,0.0038}$ & 0.298 $\pm$ 0.029   &   0.458 &  0.229  & 0         \\
\hline
2012 Oct 12 &   WHT              & NaI   &  0.1514 $\pm$ 0.0011 & 0.297 $\pm$ 0.021   &   0.421 &  0.237  & 0.0001    \\
\hline
2011 Nov 01 &   STELLA           & r'    &  0.1502 $\pm$ 0.0033 & 0.331 $\pm$ 0.045   &   0.401 &  0.227  & 0.0001    \\
2011 Nov 29 &   STELLA           & r'    &  0.1561 $\pm$ 0.0037 & 0.293 $\pm$ 0.031   &   0.401 &  0.227  & 0.0001    \\
2012 Aug 17 &   STELLA           & r'    &  0.1534 $\pm$ 0.0023 & 0.361 $\pm$ 0.024   &   0.401 &  0.227  & 0.0001    \\
2012 Oct 25 &   STELLA           & r'    &  0.1460 $^{+\,\,0.0046}_{-\,0.0048}$ & 0.315 $\pm$ 0.064   &   0.401 &  0.227  & 0.0001    \\
2012 Nov 24 &   STELLA           & r'    &  0.1518 $\pm$ 0.0042 & 0.296 $\pm$ 0.067   &   0.401 &  0.227  & 0.0001    \\
2012 Dec 22 &   STELLA           & r'    &  0.1550 $\pm$ 0.0039 & 0.274 $\pm$ 0.055   &   0.401 &  0.227  & 0.0001    \\
2013 Jan 05 &   STELLA           & r'    &  0.1537 $\pm$ 0.0022 & 0.319 $\pm$ 0.043   &   0.401 &  0.227  & 0.0001    \\
2014 Dec 21 &   Calar Alto       & Gunn r&  0.1502 $\pm$ 0.0017 & 0.310 $\pm$ 0.020   &   0.361 &  0.238  & 0.0001    \\
\hline
2011 Nov 14 &   Jena 0.6/0.9\,m  & R     &  0.1528 $\pm$ 0.0022 & 0.349 $\pm$ 0.021   &   0.387 &  0.219  & 0.0002    \\
2011 Dec 01 &   Rozhen 2.0\,m    & R     &  0.1521 $\pm$ 0.0036 & 0.377 $\pm$ 0.036   &   0.387 &  0.219  & 0.0002    \\
2011 Dec 14 &   Rozhen 0.6\,m    & R     &  0.1546 $\pm$ 0.0035 & 0.376 $\pm$ 0.014   &   0.387 &  0.219  & 0.0002    \\
2011 Dec 27 &   Rozhen 2.0\,m    & R     &  0.1546 $^{+\,\,0.0039}_{-\,0.0042}$ & 0.371 $\pm$ 0.032   &   0.387 &  0.219  & 0.0002    \\
2012 Jan 15 &   Swarthmore       & R     &  0.1554 $^{+\,\,0.0027}_{-\,0.0025}$ & 0.267 $\pm$ 0.032   &   0.387 &  0.219  & 0.0002    \\
2012 Aug 15 &   Rozhen 2.0\,m    & R     &  0.1499 $\pm$ 0.0021 & 0.315 $\pm$ 0.013   &   0.387 &  0.219  & 0.0002    \\
2012 Sep 12 &   OSN 1.5\,m       & R     &  0.1602 $\pm$ 0.0088 & 0.294 $\pm$ 0.019   &   0.387 &  0.219  & 0.0002    \\
2012 Sep 12 &   Trebur 1.2\,m    & R     &  0.1580 $\pm$ 0.0035 & 0.325 $\pm$ 0.039   &   0.387 &  0.219  & 0.0002    \\
2012 Sep 14 &   OSN 1.5\,m       & R     &  0.1523 $^{+\,\,0.0020}_{-\,0.0021}$ & 0.342 $\pm$ 0.018   &   0.387 &  0.219  & 0.0002    \\
2012 Oct 31 &   SON 0.4\,m       & R     &  0.1527 $^{+\,\,0.0034}_{-\,0.0031}$ & 0.365 $\pm$ 0.036   &   0.387 &  0.219  & 0.0002    \\
2012 Nov 22 &   OSN 1.5\,m       & R     &  0.1493 $\pm$ 0.0013 & 0.291 $\pm$ 0.010   &   0.387 &  0.219  & 0.0002    \\
2013 Aug 22 &   0.35\,m          & R     &  0.1488 $^{+\,\,0.0029}_{-\,0.0033}$ & 0.412 $\pm$ 0.033   &   0.387 &  0.219  & 0.0002    \\
2013 Sep 06 &   Rozhen 2.0\,m    & R     &  0.1508 $\pm$ 0.0029 & 0.277 $\pm$ 0.022   &   0.387 &  0.219  & 0.0002    \\
2013 Sep 06 &   Jena 0.6/0.9\,m  & R     &  0.1514 $\pm$ 0.0018 & 0.312 $\pm$ 0.016   &   0.387 &  0.219  & 0.0002    \\
2013 Sep 06 &   Torun 0.6\,m     & R     &  0.1484 $\pm$ 0.0030 & 0.278 $\pm$ 0.037   &   0.387 &  0.219  & 0.0002    \\
2013 Oct 06 &   OSN 1.5\,m       & R     &  0.1492 $\pm$ 0.0009 & 0.344 $\pm$ 0.012   &   0.387 &  0.219  & 0.0002    \\
2013 Nov 01 &   Rozhen 2.0\,m    & R     &  0.1534 $\pm$ 0.0015 & 0.309 $\pm$ 0.012   &   0.387 &  0.219  & 0.0002    \\
2013 Nov 03 &   OSN 1.5\,m       & R     &  0.1519 $^{+\,\,0.0014}_{-\,0.0016}$ & 0.276 $\pm$ 0.027   &   0.387 &  0.219  & 0.0002    \\
2013 Dec 01 &   OSN 1.5\,m       & R     &  0.1498 $\pm$ 0.0025 & 0.287 $\pm$ 0.027   &   0.387 &  0.219  & 0.0002    \\
2013 Dec 29 &   Trebur 1.2\,m    & R     &  0.1489 $\pm$ 0.0032 & 0.330 $\pm$ 0.031   &   0.387 &  0.219  & 0.0002    \\
\hline
2014 Dec 21 &   Calar Alto       & I     &  0.1514 $^{+\,\,0.0016}_{-\,0.0015}$ & 0.227 $\pm$ 0.017   &   0.306 &  0.212  & 0.0007    \\
\hline
2007 Sep 24 &   FLWO 1.2\,m      & z'    &  0.1530 $^{+\,\,0.0018}_{-\,0.0016}$ & 0.188 $\pm$ 0.021   &   0.265 &  0.211  & 0.0012    \\
2007 Oct 22 &   FLWO 1.2\,m      & z'    &  0.1481 $^{+\,\,0.0019}_{-\,0.0018}$ & 0.219 $\pm$ 0.025   &   0.265 &  0.211  & 0.0012    \\
2007 Nov 06 &   FLWO 1.2\,m      & z'    &  0.1533 $^{+\,\,0.0012}_{-\,0.0013}$ & 0.188 $\pm$ 0.017   &   0.265 &  0.211  & 0.0012    \\
2007 Nov 19 &   FLWO 1.2\,m      & z'    &  0.1523 $\pm$ 0.0015 & 0.231 $\pm$ 0.036   &   0.265 &  0.211  & 0.0012    \\

\hline                                                                                                     
\end{tabular}
\end{center}
\end{table*}

\begin{table*}
\caption{Planet-to-star radius ratio $k$ per filter with relative uncertainties. }
\label{tab_joint}
\begin{center}
\begin{tabular}{lccccl}
\hline
\hline
\noalign{\smallskip}

Filter  &    $k$  &  $u_{\mathrm{fit}}$  & $u_{\mathrm{theo}}$  & $v_{\mathrm{theo}}$  &  $\Delta k_{\mathrm{third\ light}}$ \\ 
\hline
\noalign{\smallskip}
u'           &   0.15204 $^{+\,\,0.00118}_{-\,0.00123}$ &  0.543 $\pm$ 0.023 &  0.696 &  0.112  &  0         \\
B            &   0.15338 $^{+\,\,0.00073}_{-\,0.00082}$ &  0.502 $\pm$ 0.013 &  0.583 &  0.208  &  0         \\
g' + Gunn g  &   0.15304 $^{+\,\,0.00111}_{-\,0.00101}$ &  0.448 $\pm$ 0.022 &  0.535 &  0.214  &  0         \\
V            &   0.15223 $^{+\,\,0.00123}_{-\,0.00128}$ &  0.398 $\pm$ 0.014 &  0.458 &  0.229  &  0         \\
NaI          &   0.15142 $^{+\,\,0.00108}_{-\,0.00105}$ &  0.297 $\pm$ 0.021 &  0.421 &  0.237  &  0.00006    \\      
r'+ Gunn r   &   0.15216 $^{+\,\,0.00080}_{-\,0.00078}$ &  0.317 $\pm$ 0.016 &  0.392 &  0.230  &  0.00007   \\     
R            &   0.15071 $^{+\,\,0.00053}_{-\,0.00057}$ &  0.315 $\pm$ 0.010 &  0.387 &  0.219  &  0.00021    \\    
I            &   0.15141 $^{+\,\,0.00155}_{-\,0.00147}$ &  0.227 $\pm$ 0.017 &  0.306 &  0.212  &  0.00069    \\     
z'           &   0.15149 $^{+\,\,0.00073}_{-\,0.00078}$ &  0.221 $\pm$ 0.017 &  0.265 &  0.211  &  0.00115    \\                                                                                           
                                                                                                              
\hline                                                                                                     
\end{tabular}
\end{center}
\end{table*}

\begin{table*}
\caption{Fit statistics of theoretical models against the derived transmission spectrum.}
\label{tab_chi2}
\begin{center}
\begin{tabular}{l|ccc|ccc}
\hline
\hline
\noalign{\smallskip}

  &    \multicolumn{3}{|c|}{This work}   & \multicolumn{3}{c}{This work \& Paper~I, $\lambda<720$~nm} \\
Models  &    $\chi^2$  &  N,\ DOF  & $P$  &  $\chi^2$  &  N,\ DOF  & $P$\\
\hline
\noalign{\smallskip}
Solar composition, clear           &  62.9  & 9,\ 8  & $\ll0.001$  &  219.6  &  49,\ 48  &  $\ll0.001$     \\
Solar comp., without TiO, clear    &  21.9  & 9,\ 8  & 0.005       &  106.2  &  49,\ 48  &  $\ll0.001$     \\
Rayleigh                           &  7.6   & 9,\ 8  & 0.50        &  70.9   &  49,\ 48  &  0.017     \\
Flat                               &  10.3  & 9,\ 8  & 0.24        &  43.8   &  49,\ 48  &  0.65     \\

\hline                                                                                                     
\end{tabular}
\end{center}
\end{table*}

\section{Discussion}
\label{chap_disc}
A flat spectrum of a hot Jupiter exoplanet could be explained by a thick cloud layer at high altitudes acting as a grey absorber. A comparable case of a cloudy atmosphere is the hot Jupiter WASP-31\,b with a cloud deck at only 1~mbar \citep{Sing2015}, and the super-Earth GJ1214b, whose transmission spectrum is featureless \citep{Kreidberg2014} without measurable indications for scattering at blue wavelengths \citep{deMooij2013,Nascimbeni2015}. In the presence of brightness inhomogeneities on the stellar surface, a flat spectrum could also be produced by the interplay of scattering in the planetary atmosphere and bright plage regions not occulted by the  transiting planet \citep{Oshagh2014}. However, in paper I we discuss this scenario to be very unlikely.

The results of this work indicate a flat spectrum with a spectral gradient of lower amplitude than two scale heights. Interestingly, the near-UV/optical measurements of \cite{Sing2016} made with HST/STIS show no hot Jupiter spectrum without a blueward gradient. Each of the ten investigated objects exhibit an increase in $k$ from the z' band to the u' band of at least $\sim$\,two, and up to $\sim$\,six scale heights. Therefore, HAT-P-32\,b could be the hot Jupiter with the flattest spectrum measured so far.
One way to confirm this result would be a homogeneous re-analysis of the published GEMINI, GTC, and LBT ground-based transit observations or new HST measurements at near-UV and optical wavelengths with STIS. Extrapolating from the optical spectrum, we would expect a muted or absent water feature at 1.4~$\mu$m. It would be very interesting to verify this prediction with HST/WFC3 measurements. If confirmed, HAT-P-32\,b would strengthen the tentative correlation between cloud occurrence at high altitudes and a low planetary surface gravity found by \cite{Stevenson2016}. 

\section{Conclusions}
\label{chap_concl}
HAT-P-32\,b is one of the most favourable targets for transmission spectroscopy in terms of host star brightness and predicted amplitude of the potential transmission signal. In Paper~I, we obtained a transmission spectrum from 330 to 1000~nm showing a tentative increase of the planet-star radius ratio towards blue wavelength of low amplitude. In this work, we collected the largest sample of broad-band transit photometry used for spectrophotometry so far to follow-up on this slope in the planetary spectrum. The light curves were taken in nine different filters from Sloan u' over several Johnson bands and one specific filter centred at the NaI D line to the Sloan z' band. The resulting spectrum was independent of the choice of the stellar limb darkening law in the analysis. However, the spectral gradient was dependent on the treatment of the limb darkening coefficients in the fit. We advocate the inclusion of the linear coefficient as a free parameter in the light curve modelling because it resulted in a 
better fit in all bandpasses.

While the new measurements of this work were of sufficient precision to rule out clear atmosphere models with solar metallicity, they could not distinguish between a Rayleigh-slope model and a wavelength-independent planet-star radius ratio. However, the new data helped to verify the proposed spectral gradient in showing that the $k$ values redward of 720~nm of Paper~I might be too low by about one scale height, potentially caused by systematics in the measurements. Excluding these data favoured a spectrum that is flat over the entire wavelength range. HAT-P-32\,b might be the hot Jupiter with the flattest spectrum observed so far. However, we suggest the combination of the currently available ground-based data by a homogeneous re-analysis, and the performance of transit observations with HST/STIS to confirm this peculiar result.

\section*{Acknowledgements}
This article is based on observations made with the STELLA robotic telescopes in Tenerife, an AIP facility jointly operated by AIP and IAC, the IAC80 telescope operated on the island of Tenerife by IAC in the Spanish Observatorio del Teide, the Nordic Optical Telescope, operated by the Nordic Optical Telescope Scientific Association at the Observatorio del Roque de los Muchachos, La Palma, Spain, of the IAC, the William Herschel Telescope, operated by the Isaac Newton Group and run by the Royal Greenwich Observatory at the Spanish Roque de los Muchachos Observatory in La Palma, and the 70cm telescope operated at the Babelsberg Observatory, Potsdam, Germany, by the AIP. The data presented here were obtained in part with ALFOSC, which is provided by the Instituto de Astrofisica de Andalucia (IAA) under a joint agreement with the University of Copenhagen and NOTSA. EH acknowledges support from the Spanish MINECO through grant ESP2014-57495-C2-2-R. VSD and ULTRACAM are supported by STFC. This research has made 
use of the SIMBAD data base and VizieR catalogue access tool, operated at CDS, Strasbourg, France, and of the NASA Astrophysics Data System (ADS). We thank the anonymous referee for a constructive review.





\bibliographystyle{mnras}
\bibliography{h32_bb_bib} 




\appendix

\section{Figure of the re-analysed literature light curves}

   \begin{figure*}
   \centering
   \includegraphics[height=22cm]{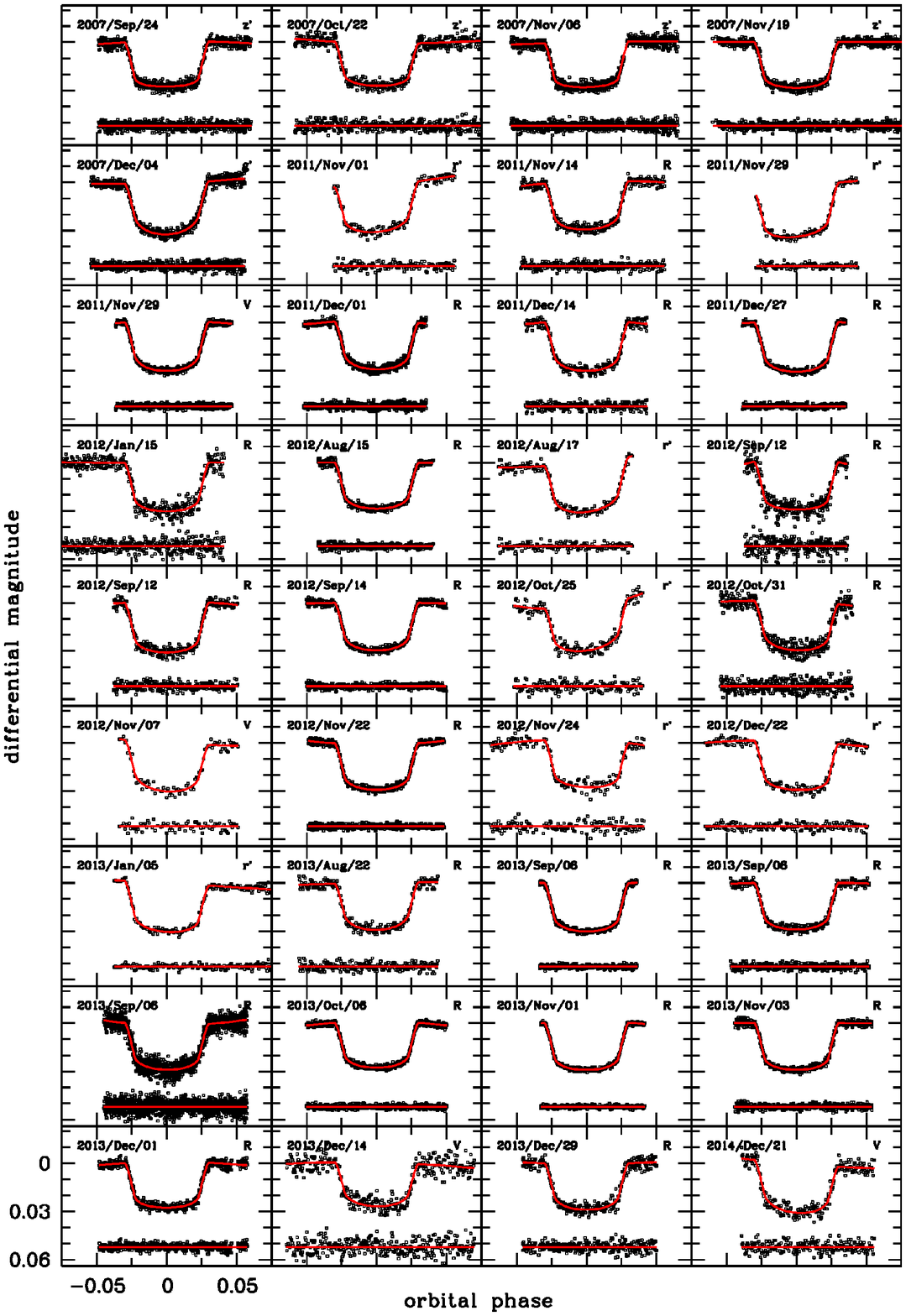}
       \caption{All literature light curves re-analysed in this work. The red solid line denotes the individual best fit model, the resulting parameter values are given in Table \ref{tab_indiv}. Below the light curves, their corresponding residuals are shown. The scale of all panels is identical with the tickmarks labeled in the lower left.}
         \label{plot_lc_lit}
   \end{figure*}



\bsp	
\label{lastpage}
\end{document}